\documentclass{conm-p-l}

\theoremstyle{definition}

\theoremstyle{remark}

\numberwithin{equation}{section}

\begin{document}

\title[Quantum field theory]{Spectral problems from quantum field theory}

\author{Dmitri V. Vassilevich}
\address{Instutut f\"{u}r Theoretische Physik, Universit\"{a}t Leipzig,
Augustusplatz 10, D-04109 Leipzig, Germany}
\address{V.~A.~Fock Institute of Physics, St.~Petersburg University, Russia}
\email{Dmitri.Vassilevich@itp.uni-leipzig.de}
\thanks{This work is supported in part by the DFG Project BO 1112/12-1.}


\subjclass[2000]{58J50, 81T20}
\date{}

\begin{abstract}
We describe how spectral functions of differential
operators appear in the quantum field theory context. We formulate
consistency conditions which should be satisfied by
the operators and by the boundary conditions. We review some modern
developments in quantum field theory and strings and show which
new spectral and boundary value problems arise.  
\end{abstract}

\maketitle
\section{Introduction}\label{sintro}
There is no sharp boundary between physics and mathematics. 
The list of topics which are considered as being parts of these
two disciplines varies in space and time. Moreover, there are topics
which belong to both. Spectral geometry is just one of the fields
where the interaction between physicists and mathematicians has
been especially fruitful. On the other hand, stylistic and linguistic
differences between traditional physical and mathematical literature
are considerable, so that some extremists could even suggest that
there is no boundary since there is a gap.

The most immediate aim of this paper is to show how the notions of
quantum field theory (QFT) can be translated to the language of
spectral theory. Also, I would like to give the reader an idea of
which structures can appear in the QFT context, which structures
are less likely, and which are forbidden on general grounds

The way spectral functions appear in quantum theory may be
illustrated by the following simple example. It is well known that
the zero-point energy (the lowest energy level) of the harmonic
oscillator with the frequency $\omega$ is $E_\omega =\frac
{\hbar}2 \omega$ ($\hbar$ is the Planck constant). 
Therefore, the ground state energy (lowest
energy) of a system of non-interacting harmonic oscillators with
eigenfrequencies $\omega_j$ is
\begin{equation}
E_0=\frac {\hbar}2 \sum_j \omega_j \,.\label{zpe}
\end{equation}
QFT is characterised by the presence of an infinite number of
degrees of freedom, i.e. it corresponds to an infinite system of
harmonic oscillators with eigenfrequencies defined by eigenvalues
of a differential operator. The sum (\ref{zpe}) is typically
divergent, but may be regularised by relating it to the zeta
function of the operator in question.

The first spectral function which appeared in QFT was not,
however, the zeta function but the heat kernel which was used by
Fock \cite{Fock37} in 1937 to represent the Green functions.
Later, in 1951, this representation was used by Schwinger
\cite{Schwinger51} in his famous work on quantum electrodynamics.
DeWitt \cite{DeWitt} made the heat kernel a standard tool to study
QFT in curved space-time. In mid 1970s Dowker and Critchley
\cite{Dowker:1975tf} and Hawking \cite{Hawking:1976ja} introduced
the zeta function regularization thus giving a precise meaning to
the idea sketched in the previous paragraph.

During the same period important developments appeared in mathematics
as well. The 1949 papers by Minakshisundaram \cite{Min49,MP49}
had much influence on the theoretical physics research. 
The Atiyah-Singer Index Theorem \cite{AP63}
found many applications in gauge theories. 
The works of Seeley \cite{Seeley68} who analysed asymptotics of the spectral
functions, and of Gilkey who suggested \cite{Gilkey75}
the most effective way for actual calculation of these asymptotics
were essential for QFT in curved space-time.

There are many books and review papers which treat the problems
of common interest for spectral geometry and quantum field 
theory\footnote{I cannot mention all publications which may seem
relevant here or in the text below. I ask the authors whose
works are omitted for understanding.}. In particular, the
monographs by Gilkey \cite{Gilkey:1995} and
Grubb \cite{Grubb:1996} contain a very detailed description of 
relevant mathematics. The book by Kirsten \cite{Kirsten:wz}
deals also with some physical applications as the Casimir energy
and the Bose-Einstein condensation. The reviews 
by Barvinsky and Vilkovisky
\cite{Barvinsky:an} and by Vassilevich \cite{Vassilevich:2003xt} 
(see also a shorter version
\cite{Vassilevich:2001wf}), as well as the book by Elizalde
\cite{Elizalde}
are oriented to the physicists.
In the present paper I introduce some new structures which
appeared recently in QFT and may be of interest for the experts
in spectral geometry. 

This paper is organised as follows. In the next section I briefly
introduce the path integral quantisation, the effective action,
 and the semiclassical expansion.
The leading order of this expansion is defined by spectral functions
of some differential operator which may be derived from the classical
action. I also discuss which properties some general properties of the
operator and of the boundary conditions. This section describes
objects instead of rigorously defining them. However, one can
give precise mathematical sense to most of the constructions
presented here. The interested reader can consult the
introduction to QFT specially tailored for the mathematicians
\cite{Deligne}. Sec.\ \ref{soneloop} discusses main spectral
functions appearing in the context of QFT. In particular, the 
divergences (``infinities'') of the effective 
action are defined by the heat trace
asymptotics, and the ``finite'' part is the zeta-determinant. 
Quantum anomalies are related to localised zeta functions.
In this section we also discuss which properties of the heat
trace asymptotics are essential to have a meaningful quantum
field theory. Throughout this paper we discuss bosonic fields
theories and Laplace type operators. Fermionic theories give
rise to operators of Dirac type. An introduction to the theory
of Dirac operators with applications to QFT can be found in the monograph
\cite{Esposito:1997mw} by Esposito. Sec.\ \ref{sexam} contains examples
of several new problems which appeared in physics over the recent
years. In particular, sec.\ \ref{sstring} is devoted to string
theory. Here we discuss which boundary conditions correspond to
open strings, and which to the Dirichlet-branes.
We also comment on string dualities and non-commutativity
(as it comes from strings). In sec.\ \ref{sbrane} we consider
domain walls and the so-called brane-world scenario. These
configurations can be described as two smooth manifolds glued
together along a common boundary. Instead of boundary conditions
one has matching conditions on the interface surface 
in this case. Last sections contain
remarks on supersymmetric theories and on non-commutative
field theories respectively.

I am grateful to Bernhelm Booss-Bavnbek, Gerd Grubb and
Krzysztof Wojciechowski for their kind invitation to
and warm hospitality at the Spectral Geometry Workshop
at Holbaek.
\section{Quantum Filed Theory and the Path Integral}\label{spath}
{\it Classical Field Theory} consists of (i) a Riemannian manifold
$M$ called the space-time, (ii) a Hermitian vector bundle $V$
with sufficiently smooth sections $\phi$ which 
are called fields, and (iii) a classical
action $S$ defined on $\phi$ with values in  $\mathbb{R}$.

Some comments are in order. Strictly speaking, the space-{\textit{time}}
should be a \textit{pseudo-}Riemannian manifold. Transitions between
the Riemannian (Euclidean) and pseudo-Riemannian (Minkowski) signatures
of the metric are performed by the so-called Wick rotation which
introduces an imaginary time coordinate (relations between spectral 
theory and the Wick rotation are discussed by Fulling \cite{Fulling}). 
In this manner most (but not all) properties of an Euclidean theory
may be translated to the Minkowski context. Besides, Euclidean field
theories contain much of important physics and interesting mathematics 
of their own. In this paper we restrict ourselves to Euclidean theories.

In many cases, classical fields do not form a vector bundle since the
fibres may have a more complicated geometry. However, in this
work we shall restrict ourselves to the perturbative analysis, i.e.
we shall work with small fluctuations about a given background field.
Such small fluctuations always form a vector space for any model.
We also assume given a Hermitian structure though it is not always
uniquely defined. It is natural to assume that the 
classical action $S$ is bounded from the below\footnote{
Euclidean gravity is a famous exception \cite{Gibbons:ac}.}.  
Local minima of the classical action are called classical solutions
or classical trajectories of the theory. 

Fundamental theories of physics are local, i.e. the action has the
form
\begin{equation}
S=\int_M \mathcal{L}_{\rm int}+\int_{\partial M} \mathcal{L}_{\rm bou}\,,
\label{locact}
\end{equation}
where $\mathcal{L}_{\rm int}$ and $\mathcal{L}_{\rm bou}$ depend on fields
at a given point and on finite number of their derivatives.

Probably the most popular example of a field theory model is the so
called $\phi^4$ theory in $n=4$ dimensions. For simplicity, we suppose
that $V$ is a real line bundle. Then the classical action reads: 
\begin{equation}
S_{\phi^4}=\int_M \left[ (\nabla\phi )^2 + m^2\phi^2 + g \phi^4 \right]
\label{phi4}
\end{equation}
Here $m^2$ and $g$ are (positive) constants. 

To quantise a given classical theory one has to replace the classical 
fields $\phi$ by operator valued distributions in a suitably defined
Hilbert space. The final aim is to be able to calculate the so-called
vacuum expectation values of arbitrary polynomials of the field operators.
For these objects there is a ``path integral'' 
representation\footnote{A simple and clean derivation 
of this representation can be found in the excellent text 
book \cite{Faddeev:be} by Faddeev and Slavnov.}:
\begin{equation}
\langle \phi (x_1) \dots \phi (x_k) \rangle_0 =
\frac 1N \int (\mathcal{D}\phi)\, \phi (x_1) 
\dots \phi (x_k) e^{-\frac 1\hbar S}\,.
\label{pint}
\end{equation}
Here the bracket $\langle \dots \rangle_0$ denotes the vacuum expectation
value, $\hbar$ is the Planck constant, $(\mathcal{D}\phi)$ is the integration
measure, $1/N$ is a normalisation constant chosen in such a way that
$\langle 1 \rangle_0=1$. 

Let me stress that the vacuum expectation values of field polynomials
contain practically all information which is needed in QFT. The
representation (\ref{pint}) has to be derived from basic principles
of quantum mechanics, but it can be easily understood on its' own.
The right hand side of (\ref{pint}) is nothing else than a statistical
average with the weight $e^{-\frac 1\hbar S}$. In the formal limit
$\hbar\to 0$ the path integral is dominated by the minima of
the classical action $S$ thus recovering the classical theory.

The integral (\ref{pint}) is infinite-dimensional and, as it stays,
is ill-defined. Besides, the measure $(\mathcal{D}\phi)$ may have
a very complicated structure. We shall ignore these difficulties
in this section and just deal with the path integral as with an
ordinary integral. This will be enough to achieve a qualitative understanding
of what is going on in quantum field theory.  

The vacuum expectation values can be generated by taking repeated
functional derivatives of the functional
\begin{equation} 
Z(J) = \frac 1N \int (\mathcal{D}\phi)\, e^{-\frac 1\hbar (S +\int_M J\phi ) }
\label{genfunc}
\end{equation}
with respect to the ``external source'' $J$.

There is another functional, $W(\bar\phi)$, 
which is given by the Legendre transform of
$-\ln Z(J)$. It is called the effective action and is defined by the
equation
\begin{equation}
e^{-\frac 1\hbar W(\bar\phi )}=
\frac 1N \int (\mathcal{D}\phi)\, e^{-\frac 1\hbar \left( S(\bar\phi +\phi )+
\int_M J \phi \right) } \,,\label{effact}
\end{equation}
where $J$ is not an independent variable any more. It should be
expressed in terms of the {\it background field} $\bar \phi$ by means of
\begin{equation}
\frac{\delta W(\bar\phi)}{\delta \bar\phi} = -J\,.
\label{Jeq}
\end{equation} 

The effective action $W$ contains the same information as $Z(J)$ but
is somewhat easier to analyse. Let us consider a {\it semiclassical
expansion} of $W$. This means $\hbar \to 0$ asymptotics of the
equations (\ref{effact}) and (\ref{Jeq}). To this end one has to
use the saddle point method to evaluate the integral in (\ref{effact}).
Let us expand the classical action $S(\bar\phi + \phi)$ about $\bar\phi$,
\begin{equation}
S(\bar\phi + \phi )=S(\bar\phi )+ \int \frac{\delta S}{\delta \bar\phi (x)}
\phi (x) + \frac 12
\int \int \frac{\delta^2 S}{\delta \bar\phi (x) \delta\bar\phi (y)}
\phi (x) \phi (y) + O(\phi^3) \,.\label{expS}
\end{equation}
the zeroth order approximation to the effective action $W$ is just
the classical action:
\begin{equation}
W_0(\bar\phi )=S(\bar\phi ) \,.\label{zeroW}
\end{equation}
Consequently, to this order 
\begin{equation}
J_0=-\frac{\delta S(\bar \phi )}{\delta\bar\phi} \,,
\label{zeroJ}
\end{equation}
so that the linear terms in $\phi$ in the exponential (\ref{effact})
are cancelled. The next approximation to the effective action
is obtained by keeping the quadratic term in (\ref{expS}) and
performing the Gaussian integration. Usually there exists a
(pseudo)differential operator $D$ such that
\begin{equation}
S_2(\bar\phi ,\phi):=\int_x \int_y \frac{\delta^2 S(\bar \phi)}{
\delta \bar\phi (x) \delta\bar\phi (y)}
\phi (x) \phi (y) = \int \phi D[\bar\phi ] \phi \,.\label{SD}
\end{equation}
Therefore,
\begin{equation}
W_1(\bar\phi )=\frac {\hbar}2 \ln \det (D) \label{lndetD}
\end{equation}
We stress again that we are working with {\it bosonic} theories
only. For fermions the rules of functional integration are considerably
different, so that (\ref{lndetD}) is no longer true.

This semiclassical expansion is also called the loop expansion
since in the language of Feynman diagrams $W_q$ is described by
graphs containing $q$ loops.

One can also define the correlation functions of $\phi$ in the
presence of the background field $\bar \phi$. In this case one
has to keep the sources unrestricted to be able to vary with respect
to them. Then the semiclassical expansion looks a little bit more
complicated. In the leading order in $\hbar$ the result reads:
\begin{equation}
\label{phicor}
\langle \phi (x)  \phi (y) \rangle \sim \mathcal{G}(x,y)\,,
\end{equation}
where $\mathcal{G}(x,y)$ is the Green function, $D\mathcal{G}=Id$.

The operator $D$ should satisfy some natural restrictions:
\begin{enumerate}
\item The operator $D$ should be symmetric. Otherwise, the Gaussian
integral would not produce $\det (D)$.
\item Since the fundamental actions are local (an exception
will be discussed in sec.\ \ref{snon} below), the operator
$D$ is a partial differential operator rather than a pseudodifferential
one.
\item The operator $D$ should have finite number of negative and zero
modes. Path integration in the directions corresponding to negative
and zero modes cannot be performed by the saddle point method. These
directions should be treated separately.
\end{enumerate}

Note, that infinite number of zero modes is a characteristic feature
of {\it gauge theories} (cf. a mathematical introduction  \cite{Marathe}
by Marathe and Martucci).
During the quantisation these zero modes are removed by a gauge
fixing procedure \cite{Faddeev:be}, so that the resulting quantum
theory satisfies the restriction given above.

If $M$ has a boundary, r.h.s. of (\ref{SD}) is supplemented by
a boundary term,
\begin{equation}
S_2(\bar\phi ,\phi)=\int_M \phi D[\bar\phi ] \phi +
\int_{\partial M} \mathcal{L}^2_{\rm bou} (\bar\phi,\phi)\,,
\label{bouterm}
\end{equation}
where the boundary density
$\mathcal{L}^2_{\rm bou} (\bar\phi,\phi)$ is quadratic in $\phi$.
One has to impose some boundary conditions on $\phi$. They should
be such that (i) the properties (1) and (3) listed above hold, and
(ii) the boundary term in (\ref{bouterm}) vanishes, so that the Gaussian
integration produces eigenvalues of $D$ at least formally.

Let us now consider the example (\ref{phi4}). Obviously, the
quadratic part of the action reads:
\begin{equation}
S_2(\bar\phi ,\phi)=\int_M \phi ( -\nabla^2 + m^2 +6g\bar\phi^2 )\phi
-\int_{\partial M} \phi \nabla_n \phi \,,\label{S2p4}
\end{equation}
where $\nabla_n$ is a derivative with respect to inward
pointing unit vector on the boundary. Clearly, in this case
\begin{equation}
D= -\nabla^2 + m^2 +6g\bar\phi^2 \label{Dphi4}
\end{equation}
is an elliptic partial differential operator. Natural boundary
conditions which ensure vanishing of the boundary term in (\ref{S2p4})
are either Dirichlet
\begin{equation}
\phi \vert_{\partial M}=0 \,,\label{Dirbc}
\end{equation}
or Neumann
\begin{equation}
\nabla_n \phi \vert_{\partial M}=0 \,,\label{Neubc}
\end{equation}
ones. Both guarantee strong ellipticity of the boundary value problem
and satisfy all consistency conditions listed above. A somewhat less
trivial fact is that these conditions are also satisfied by rather
complicated mixtures of Dirichlet and Neumann conditions (cf. sec.\
\ref{sexam}).
\section{Spectral Functions and QFT}\label{soneloop}
As we have seen in the previous section, first non-trivial
quantum correction to classical action is defined by $\det (D)$.
This quantity has to be regularised. 
Let us define the {\it heat trace} $K(t,D)$ by the equation
\begin{equation}
K(t,D)={\rm Tr}_{L^2} \left( e^{-tD} \right) \,.\label{hkdef}
\end{equation}
There exist a useful
representation for the determinant
\begin{equation}
\ln \det (D) = - \int\limits_0^\infty \frac{dt}t K(t,D)\,, \label{intdet}
\end{equation}
which is still divergent, but is nevertheless useful to discuss
regularizations. One possible way to make sense of (\ref{intdet}) is to
shift the power of $t$ (and introduce a constant $\tilde\mu$
of the dimensions of mass to keep proper dimension of the
effective action) \cite{Dowker:1975tf,Hawking:1976ja} 
\begin{equation}
\ln \det (D)_s = - \tilde \mu^{2s}
\int\limits_0^\infty \frac{dt}{t^{1-s}} K(t,D)\,, \label{dets}
\end{equation}
so that the integral converges for ``typical'' $D$ for sufficiently
large $\mathrm{Re} s$. The zeta function is defined as
\begin{equation}
\zeta (s,D):={\rm Tr}_{L^2} (D^{-s}) \label{defzeta}
\end{equation}
We assume that $D$ is positive (negative and zero modes should be
excluded, cf. previous section). The heat trace is then
expressed as
\begin{equation}
K(t,D)=\frac 1{2\pi i}\int_{\mathrm{Re} s=c} t^{-s} \Gamma (s) \zeta (s,D) \,.
\label{hkzeta}
\end{equation}
$c$ should be sufficiently large.

We shall also need a ``localised'' (or ``smeared'') version of the
heat trace. Let $f$ be a smooth function on $M$ (or an endomorphism
of the vector bundle $V$, depending on the context). Then
\begin{equation}
K(f;t,D)={\rm Tr}_{L^2} \left(f e^{-tD} \right) \,.\label{hksm}
\end{equation}
Obviously, $K(t,D)=K(1;t,D)$. One can also define a localised version of
the zeta function, so that the relations between the heat trace and
the zeta function will hold also in the local sense.

We assume that the following asymptotic exists for $t\to +0$:
\begin{equation}
K(f;t,D)\simeq t^{-n/2} \left( \sum_{k=0}^{N} t^{k/2} a_k(f,D)
+ \sum_{j=N+1}^\infty t^{j/2} ( a'_{j}(f,D) \ln t +
a''_j (f,D))
\right), \label{asex} \end{equation}
so that we have mixed power and power-logarithm asymptotic
expansion. The coefficients $a_k$ are locally computable,
i.e. they can be represented as integrals of local invariants
constructed from the symbol of $D$.
This expansion indeed exists if $D$ is an elliptic second order operator
and if the boundary conditions satisfy some additional requirements
formulated by Grubb and Seeley \cite{GS95,G99}.
In the particular case when $D$ is of Laplace type and the boundary
conditions are local,
logarithms are absent ($N=\infty$). If, moreover, $M$ has no boundary,
even numbered coefficients vanish. 

For any operator of Laplace type there exist a unique connection
$\nabla$ and a unique endomorphism $E$ such that
\begin{equation}
D= -(g^{\mu\nu}\nabla_\mu\nabla_\nu + E) \label{LapD}
\end{equation}
($g^{\mu\nu}$ is Riemannian metric on $M$). Let 
$\partial M=\emptyset$. Let $R_{\mu\nu\rho\sigma}$ be the Riemann
curvature tensor, let $R_{\mu\nu}:={R^\sigma}_{\mu\nu\sigma}$
be the Ricci tensor, and let $R:=R_\mu^\mu$ be the scalar curvature.
We define the field strength of $\nabla_\mu =\partial_\mu +\omega_\mu$
by the equation $\Omega_{\mu\nu}:=\partial_\mu \omega_\nu
-\partial_\nu \omega_\nu +\omega_\mu \omega_\nu - \omega_\nu \omega_\mu$.
Then
 \begin{eqnarray}
&&a_{0}(f,D)=(4\pi)^{-n/2}\int_M {\rm tr}\{f\}. \label{a0a4}\\
&&a_2(f,D)
         =(4\pi)^{-n/2}6^{-1}\int_M {\rm tr}\{f(6E+R)\}.\nonumber \\
&&   a_4(f,D)=(4\pi)^{-n/2}360^{-1}\int_M {\rm tr}\{f(60\nabla^2 E
    +60R E+180E^2 \nonumber \\
&&\qquad\qquad
    +12\nabla^2 R+5R^2-2 R_{\mu\nu}R^{\mu\nu}
    +2 R^{\mu\nu\rho\sigma}R_{\mu\nu\rho\sigma}
    +30\Omega^{\mu\nu}\Omega_{\mu\nu})\}.\nonumber
\end{eqnarray}
These expressions appeared in both mathematical \cite{MS}
and physical \cite{DeWitt} literature.
The coefficient $a_6$ was first calculated by Gilkey \cite{Gilkey75},
$a_8$ was obtained by Amsterdamski et al\cite{Ams}
and by Avramidi \cite{Ava8}, and $a_{10}$ was calculated by van de Ven
\cite{vande}. In the presence of boundaries the calculations are much
more involved. For local mixed boundary conditions $a_4$ was calculated
by Branson and Gilkey \cite{BG90} 
(with minor corrections by Vassilevich \cite{Vas94}), and 
$a_5$ was done by Branson et al \cite{BGKV99}.

Equation (\ref{hkzeta}) implies that there is a relation between the
poles of $\Gamma (s)\zeta (s,D)$ and the asymptotic expansion for the
heat trace. In particular, if $a'_n(D):=a'(1,D)=0$ the zeta function
is regular at $s=0$. This is achieved in the case $N>n$ when also the
following important relation holds 
\begin{equation}
\zeta (0,D)=a_n(D) \,,\label{z0an}
\end{equation}
which tells that $\zeta (0,D)$ is locally computable.

For sufficiently large $\mathrm{Re} s$ the integral (\ref{dets}) yields
\begin{equation}
\ln \det (D)_s = - \tilde \mu^{2s} \Gamma (s) \zeta (s,D) \,.
\label{Gaz}
\end{equation}
Now the right hand side has to be analytically continued
to the physical value $s=0$. If $\zeta (s,D)$ is regular near
$s=0$ (i.e. if $a'_n(D)=0$)
the only singularity of (\ref{Gaz}) at this point comes from
a pole of the gamma function. One can expand near $s=0$:
\begin{equation}
\ln \det (D)_s \simeq -\left( \frac 1s -\gamma_E +\ln \tilde\mu^2 \right)
\zeta (0,D) - \zeta'(0,D)\,, \label{detpole}
\end{equation}
where $\gamma_E$ is the Euler constant. The second term on the right
hand side of (\ref{detpole}) is nothing else than the Ray-Singer
\cite{RS71} definition of the determinant.

One can see that the r.h.s. of (\ref{detpole}) is divergent as $s\to 0$,
and therefore the one-loop effective action (\ref{lndetD}) is divergent
as well. However, one can ensure that the sum of the zero-loop part
(\ref{zeroW}) and of the one-loop part is convergent in the limit
$s\to 0$. Let the classical action $S$ depend on fields $\phi$ and
``charges'' $e_j$. One can replace in $S$: $\phi\to Z_\phi\phi$,
$e_j\to Z_j e_j$ (no summation over $j$) with $Z_{j,\phi}=
1+\hbar z_{j,\phi} +O(\hbar^2)$. The constants $z$ can depend on
charges and on the regularization parameter $s$ but cannot depend
on $\phi$. Obviously, the classical limit $\hbar =0$ remains
unchanged. One can try to find such values of $z$ that the
sum $S(\bar\phi)+W_1(\bar\phi )$ becomes regular at $s=0$.
If this is possible to do by introducing a {\it finite}
number of renormalization constants $Z_j$ the theory is called
multiplicatively renormalizable (at one loop). One can give a precise
mathematical meaning to the renormalization procedure. Here we shall
need just one simple observation. In a renormalizable theory
the divergent part of the one-loop effective action should repeat
the structure of the classical action. In particular, the divergent
part must be {\it local}.

In order to see how this procedure works let us consider the 
$\phi^4$ theory (\ref{phi4}) on a compact flat 4-dimensional
manifold without boundary\footnote{This is, of course,
an oversimplified example.}. Relevant operator is given by
(\ref{Dphi4}), it is of Laplace type, $E=-m^2-6g^2\bar\phi^2$. Then, by 
using (\ref{a0a4}) and (\ref{z0an}), one obtains
\begin{equation}
\zeta (0,D)=a_4=\frac 1{32\pi^2} \int_M [ m^4 +12gm^2\bar\phi^2
+36 g^2 \bar\phi^4]\,.\label{p4z0}
\end{equation}
By (\ref{lndetD}) this expression defines the one-loop divergences.
We see, that the pole part of the effective action indeed repeats
the form of the classical action (\ref{phi4}) up to a field independent
term $m^4$ which can be neglected (unless we treat metric as a
dynamical field, which is a much more complicated case). The remaining
divergences can be removed by choosing
\begin{equation}
z_{m^2}=\frac{3g}{16\pi^2 s} \,,\qquad z_g=\frac {9g}{16\pi^2s} 
\,,\label{renp4}
\end{equation}
so that the $\phi^4$ theory is indeed multiplicatively renormalizable
(at least at the one-loop level).

We have seen that the properties of the heat trace expansion (\ref{asex})
for $f=1$ are important for the renormalization theory. The localised
heat trace asymptotics ($f\ne 1$) are also very useful. Probably the
most spectacular application is quantum {\it anomalies}. Consider
a family of the operators
\begin{equation}
D_{[\alpha \rho]}=e^{\alpha\rho} D e^{\alpha\rho}\,,\label{fam}
\end{equation}
where $\rho$ is an endomorphism of $V$, and $\alpha$ is a number.
Then 
\begin{equation}
\frac{d}{d\alpha} \zeta (s,D_{[\alpha \rho]})=
-2s {\rm Tr} (\rho D_{[\alpha \rho]}^{-s})
=-2s \zeta (\rho, s,D_{[\alpha \rho]}) \,.
\label{trafz}
\end{equation}
By comparing this equation to (\ref{detpole}) one sees that
if the localised zeta function $\zeta (\rho, s,D_{[\alpha \rho]})$
is regular and locally computable at $s=0$, the derivative of 
$\log \det (D_{[\alpha \rho]})$ with respect to $\alpha$
is finite for $s\to 0$ and local. In other words, one can take
the $s\to 0$ limit in (\ref{detpole}) to write:
\begin{equation}
\frac{d}{d\alpha} \ln \det (D_{[\alpha \rho]})=
2\zeta (\rho, 0,D_{[\alpha \rho]}) \,.\label{anom}
\end{equation}

For a hermitian $\rho$
the transformation $D\to D_{[\alpha\rho]}$ can be made a symmetry of
the quadratic part of the classical action (\ref{SD}) 
if accompanied by the transformation
$\phi \to e^{-\alpha \rho}\phi$ of fluctuations. 
If it can be promoted  to
a symmetry of full classical action, then r.h.s.\ of (\ref{anom})
is called quantum anomaly since it describes non-invariance
of the quantum action with respect to the same transformations.

Given explicit form of the zeta function in (\ref{anom}) this
equation can be integrated to give $\ln \det (D_{[\rho]}) -
\ln \det (D_{[0]})$. This is especially important if $D_{[0]}$
is trivial in some sense. Then one can obtain the effective
action (determinant) itself. Probably the most famous
example of such construction is the Polyakov effective
action \cite{Pol} which is obtained by integration of
the conformal anomaly on a two dimensional manifold (i.e.
when $\rho$ is a real function).
A more recent example is duality symmetries of the $p$-form
theories considered by Gilkey et al \cite{Gilkey:2002qd}.

In general, the analysis of a quantum theory at the one-loop 
approximation (by spectral theory methods) consists of the
following steps. One starts with defining the operator $D$
and with fixing an appropriate set of boundary conditions
(cf. sec.\ \ref{spath}). Then one has to make sure that
an expansion of the type (\ref{asex}) exists. Then, if
the $\log$-term $a'_n$ is absent and if $a_n=a''_n$
is local, one can analyse renormalization in the usual
way and calculate the anomalies. Note, that strictly speaking
it is not excluded on general grounds that the theory can be
renormalised even if $a'_n\ne 0$. However, no example
of such theory is known. As soon as the renormalization is
done, one can try to calculate finite part of the effective
action which is essentially defined by $\zeta'(0,D)$ (cf.
eq.\ (\ref{detpole})). There exist very few examples of the
theories where this can be done in a closed analytical form.
Therefore, one is usually restricted to various expansions
of the effective action (cf. \cite{Vassilevich:2003xt}
for an overview).
\section{Examples}\label{sexam}
During its' earlier history Quantum Field Theory dealt with
Laplace type operators either on manifolds without boundaries
or with simplest (Dirichlet or Neumann) boundary conditions.
Recent new developments introduced new geometries and new
spectral problems to physics. Several examples will be considered
in this section. 
\subsection{Strings}\label{sstring}
String \cite{Polchinski:rq}
is a one-dimensional object moving in a $d$-dimensional
manifold called the target space. During this movement
a two-dimensional submanifold $M$ called the world surface is formed.
We denote by $X^j$ a local coordinate system in the target space, and  
by $x^\mu$ a coordinate system on the world surface. Dynamics 
of string is defined by the embedding functions $X^j(x)$.
We assume that the target space is equipped with a Riemannian
metric $G_{ij}(X)$ and with some other fields $B_{ij}(X)$,
$A_j(X)$, etc. Then propagation of the string is described by the
world-surface action 
\begin{eqnarray}&&
S^{[\sigma]}=\int_M d^2 x\left(
\sqrt{g} G_{ij}(X) g^{\mu\nu}
\partial_\mu X^i \partial_\nu X^j +
\epsilon^{\mu\nu} B_{ij}(X) \partial_\mu X^i \partial_\nu X^j \right)
\nonumber\\
&&\qquad\qquad +\int_{\partial M} A_jdX^j\,.
\label{actsigma}
\end{eqnarray}
Here $g^{\mu\nu}$ and $\epsilon^{\mu\nu}$ are the metric tensor and
the Levi-Civita tensor on the world surface respectively.
We have absorbed the string tension $\alpha'$ (which usually appears in
the action) in a field redefinition.
$A^j$ plays a role of the electromagnetic potential. We see, that
electric charges of the string are concentrated at the end points.

We are dealing with a two-dimensional field theory where the coordinates
of the string $X^j(x)$ play the role of fields (the same 
as $\phi$ in the sections
above). The quantities $G$, $B$ and $A$ (which are fields on the target
space) play the role of charges (or couplings) in this two-dimensional
theory. Since  $G$, $B$ and $A$ are almost arbitrary functions of $X$
we have a system with infinitely many charges (each charge corresponds
to a power of $X$ in Taylor series expansions of $G$, $B$ and $A$). 

To apply the background field formalism one has to fix a trajectory
of the string $\bar X(x)$ and consider small deviations $\xi$ from
this trajectory, $X=\bar X+\xi$ ($\xi$ then belongs
to the tangent bundle of the target space manifold restricted
to the string world surface). Then one has to expand the action
in a power series of $\xi$ up to the quadratic order. With increasing
level of generality this was done e.g. by Braaten, Curtright and Zachos
\cite{Braaten:is}, Osborn \cite{Osborn:gm},
Kummer and Vassilevich \cite{Kummer:2000ae}. Here we are interested in
the boundary term only:
\begin{equation}
S_2^{\rm bou}=-\frac 12 \int_{\partial M} d\tau \xi
\left( \nabla_n + \frac 12 \left( \nabla_\tau \Gamma
+\Gamma \nabla_\tau \right) + \mathcal{S} \right) \xi \,,\label{boustr}
\end{equation}
where $\nabla$ is a connection, $\Gamma$ and $\mathcal{S}$ are
some endomorphisms which depend on $A(\bar X)$, $G(\bar X)$,
$B(\bar X)$.

There exists a natural choice of the boundary conditions
\begin{equation}
\left( \nabla_n + \frac 12 \left( \nabla_\tau \Gamma
+\Gamma \nabla_\tau \right) + \mathcal{S} \right) \xi 
\vert_{\partial M}=0\,,\label{openstr}
\end{equation}
for which the boundary action (\ref{boustr}) vanishes and the operator
$D$ (not presented here explicitly) is symmetric. These boundary conditions
describe ``free'' propagation of an open string in the target
space manifold. This is 
probably the most important case in physics when the boundary
conditions contain both normal and tangential derivatives\footnote{
Other examples of such boundary conditions include gravity
\cite{Marachevsky:1995dr,Avramidi:1996ae} and some solid state systems
\cite{Bordag:1999ux}.}. Such boundary conditions (called
{\it oblique}) appeared also in the mathematical papers by Grubb
\cite{Grubb74} and by Gilkey and Smith \cite{GS}. 
The study of the heat trace asymptotics was initiated
by McAvity and Osborn \cite{McAvity:xf} and then continued by
Dowker and Kirsten \cite{Dowker:1997mn,Dowker:1998hm}.
Avramidi and Esposito \cite{Avramidi:1997sb,Avramidi:1998xj}
lifted some commutativity assumptions and
proved a simple criterion of strong ellipticity.

In contrast to Dirichlet and Neumann boundary value problems oblique
boundary conditions are strongly elliptic only if $|\Gamma |$ is
sufficiently small. For large $|\Gamma |$ the operator $D$
has infinitely many negative modes, so that the heat kernel is
not of trace class any more. Physically, lack of strong ellipticity
means that the string endpoints are forced to move faster than light,
so that the system develops instabilities. If strong ellipticity is
preserved, there exists the asymptotic expansion (\ref{asex})
without log-terms, so that all coefficients are locally computable.
One can then easily check that the model is renormalizable
and calculate the counterterms. The condition that these counterterms
vanish is very important in string theory since it is equivalent
to the equations of motion of a (super-) gravity theory for the
fields $G$, $A$, $B$ on the target space manifold. This is probably
the most straightforward way to derive the low-energy limit of
the string theory.

The condition (\ref{openstr}) is not the only possible choice.
Let us assume given two local complementary projectors $\Pi_+$
and $\Pi_-$. Then we may impose the conditions (\ref{openstr})
on $\Pi_+\xi$ and Dirichlet boundary conditions on $\Pi_-\xi$.
Such configurations, introduced in the string theory context
by Dai, Leigh and Polchinski  \cite{Dai:ua},
are called Dirichlet branes (or D-branes).
Physically these boundary conditions mean that the string
endpoints are confined in a submanifold of the target space.
Spectral properties (strong ellipticity, absence of logarithms,
locality of the counterterm) of D-branes are very similar to that
of the open strings. Therefore, we shall not consider
them here in detail.

A very interesting idea of string theory which has not been fully
explored yet from the spectral theory point of view is the string
dualities. It has been observed \cite{Dai:ua,Horava:1989ga,Green:et} 
that by
suitably transforming the target space fields $A$, $B$, $G$ and by
interchanging oblique (open string) and Dirichlet (D-brane)
boundary conditions (i.e. by exchanging the role of $\Pi_-$ and
$\Pi_+$) one arrives at a quantum theory which is equivalent to
the initial one. Such transformation is called target space
duality (or T-duality). An important property of the duality
transformations is that they map "strong coupling" regimes to
"weak coupling" regimes \footnote{One can keep in mind the example
of ordinary electrodynamics. The duality transformation, which is
the Hodge duality of the electromagnetic field strength,
interchanges electric and magnetic fields, and also interchanges
electrically charged particles with the charge $e$ with
magnetically charge particles (monopoles) with the magnetic charge
$\sim 1/e$. }. In some simple cases, as explained by Schwarz and Tseytlin
\cite{Schwarz:1992te} and by Vassilevich and Zelnikov 
\cite{Vassilevich:2000kt},
the duality symmetry leads to equivalence of determinants of some
(non-isospectral) operators. Could it be that in this way one may
obtain more interesting (and yet overlooked) relations between
determinants?

Another important feature of string theory is that it
leads to a non-commutative geometry on the target space. To illustrate this
point let us consider a rather simple particular case of the string
dynamics when the world surface of the string is 
$M=\mathbb{R}\times \mathbb{R}_+$, the target space is $\mathbb{R}^d$
with standard flat metric $G_{ij}=\delta_{ij}$, with zero electromagnetic field
$A$ and with a constant field $B$. In this case 
$D=-\partial_1^2 -\partial_2^2$, and the boundary condition 
(\ref{openstr}) becomes:
\begin{equation}
\left( \partial_2 \delta_{jk}+ B_{jk}\partial_1 \right) \xi^k 
\vert_{\partial M}=0 \,,\label{simplebc}
\end{equation}
where $\partial_2$ is a 
normal derivative, and $\partial_1$ is a tangential one.
This problem can be explicitly solved. In particular, one can find
the Green function $\mathcal{G} (x,y)$. When both point $x$ and $y$
are on the boundary it reads  
\cite{Schomerus:1999ug,Seiberg:1999vs}
($\tau :=x^1,\ \tau':=y^1$):
\begin{equation}
\mathcal{G}(\tau,\tau')=-C \log (\tau -\tau')^2 +\frac 12 i \theta
{\rm sign} (\tau -\tau') \label{Npro}
\end{equation}
where
\begin{equation}
C=(1+B^2)^{-1},\qquad \theta = iB (1+B^2)^{-1} \label{Dtheta}
\end{equation}
and all multiplications are $d\times d$ matrix multiplications.

Next we need further input from Quantum Field Theory, which we use here
without going into the derivation. One has to interpret $\tau$ as a time
coordinate\footnote{This means that one has to consider the theory in
a pseudoeuclidean space. In such a case the field $B$ has to be
replaced by $iB$. This property may be considered as just another
miracle of quantum theory, but it is important if one 
compares the formulae below to other results in the literature}.
The commutators of the $\xi$ is then related to time-ordered
correlators:
\begin{equation}
[\xi^j(\tau ),\xi^k (\tau )]=\lim_{\tau' \to \tau}
\langle   \xi^j (\tau_> )\xi^k(\tau_< )-
\xi^k (\tau_> )\xi^j(\tau_< ) \rangle \,,\label{commut}
\end{equation} 
where $\tau_>$ (resp.\ $\tau_<$) is larger (smaller) of the two
arguments $\tau$ and $\tau'$. Now we can relate the correlators to the Green
function by means of (\ref{phicor}):
\begin{equation}
[\xi^j(\tau ),\xi^k (\tau )] = i\theta^{jk} \,.\label{nccoor}
\end{equation}
Since $\xi$ is a coordinate of the string endpoint, the equation
(\ref{nccoor}) implies that the coordinates on the target space  
do not commute. Although the arguments given above are rather incomplete
(cf. papers by Schomerus \cite{Schomerus:1999ug} and by
Seiberg and Witten \cite{Seiberg:1999vs} where I have borrowed
these arguments for more details)
there are two important lessons one can learn from them. First,
the non-commutativity structure if fully defined by the Green function
of the operator $D$, so that it can be analysed by the spectral theory
methods. Second, this effect appears anytime when the Green function
has a part which is antisymmetric in the coordinates, so that this feature
should be rather common. 

Another question which arises in the context of strings is could we
define a more general boundary conditions than the ones already considered
above? Clearly, if we demand that the boundary conditions are local,
we do not have much additional opportunities. There is no necessity,
however, to demand locality. Physically, boundary conditions
describe interactions with some states leaving exclusively on the boundary
(called {\em edge} or {\it boundary} states, such states appear
also in condensed mater problems). Especially in the view of the duality
symmetry it is clear that such states may be non-local (cf. the example
above involving electric charges and monopoles: in normal electrodynamics
electric charges are local point-like objects, while magnetic monopoles
are soliton-like extended objects). These arguments motivated Vassilevich
to introduce {\it spectral branes} \cite{Vassilevich:2001at}, i.e. to
replace the local projectors $\Pi_\pm$ by spectral projectors of the
Atiyah-Patodi-Singer \cite{Atiyah:jf} type. There is an important difference
for the APS scheme: for the bosonic string spectral boundary conditions 
have to be used for a second order operator of Laplace type, while 
the original APS proposal refers to a Dirac operator. Fortunately for
the spectral branes it has been demonstrated very recently by Grubb
\cite{Grubb:2003yr} that for the case in question there is an asymptotic
series (\ref{asex}) with vanishing leading logarithm, $a'_n=0$.
Therefore, one can indeed construct a well posed quantum field theory
(at least in the one-loop approximation) and define such important
quantities as quantum anomalies. Actual calculations of the divergences
and anomalies is still an open subject.
\subsection{Domain walls and the brane-world scenario}\label{sbrane}
Decomposition of manifolds was one of the main topics of this workshop.
Over the recent years many exciting results on spectral invariants
were obtained in this framework (see other contributions to this volume
and lectures by Park and Wojciechowski \cite{PW02}). 
Somewhat similar constructions appeared recently
in theoretical physics in the context of domain walls and of the brane
world scenario.

Sharp boundaries are not always good models for real physical systems.
A narrow potential barrier is a much better approximation in many
cases. Consider a manifold $M$ and a submanifold $\Sigma$ of the
dimension $n-1$. Let
\begin{equation}
D[v ]=D+v \delta_\Sigma \,.\label{singop}
\end{equation}
$D$ is an operator of Laplace type. 
Let $h$ be
the determinant of the induced metric on $\Sigma$. Then $\delta_\Sigma$ is a
delta function defined such that
\begin{equation}
\int_M dx\sqrt{g} \delta_\Sigma f(x) =\int_\Sigma dx \sqrt{h}
f(x) \,.\label{deltaSig}
\end{equation}
The spectral problem
for $D[v ]$ on $M$ as it stands is ill-defined 
owing to the discontinuities (or singularities) on $\Sigma$.
It should be replaced by a pair of spectral
problems on the two sides
$M^\pm$ of
$\Sigma$ together with suitable matching conditions on $\Sigma$.
Let $e_n$ be a unit normal to $\Sigma$ and let $x^n=0$ on $\Sigma$
Then the matching conditions read:
\begin{eqnarray}
&&\phi\vert_{x^n=+0}=\phi\vert_{x^n=-0}\label{mdelta}\\
&&-\nabla_n \phi \vert_{x^n=+0}
+\nabla_n \phi \vert_{x^n=-0} +v\phi =0\,.\nonumber
\end{eqnarray}
The first line (continuity of $\phi$) is required to make sense
of the multiplication of $\phi$ by a delta-function. The second
line can be ``derived'' by considering the eigenvalue equation for
$D[v]$ in a vicinity of $\Sigma$. 

Further generalisations of this constructions are suggested by
the so-called brane-world scenario of Randall and Sundrum
\cite{Randall:1999ee,Randall:1999vf} which became a hot topic
in theoretical physics a few years ago. According to this
scenario our world
is a four dimensional membrane in a five dimensional space
(a similar proposal was made earlier by Rubakov and
Shaposhnikov \cite{Rubakov:bb}). Typical form of the metric near 
$\Sigma$ is 
\begin{equation}
(ds)^2=(dx^n)^2 + e^{-\alpha |x^n|} (ds_{n-1})^2, \label{branemetr}
\end{equation}
where $\alpha$ is a constant and where $(ds_{n-1})$ is a line element
on the $(n-1)$-dimensional hypersurface $\Sigma$. Due to the presence of the
absolute value of the $n$-th coordinate in (\ref{branemetr}), the normal
derivative of the metric jumps on $\Sigma$.
One can think of two smooth manifolds
$M^+$ and $M^-$ glued together along their common boundary $\Sigma$.
Neither Riemann tensor, nor matrix potential $E$ must be continuous
on $\Sigma$. Also, the extrinsic curvatures $L_{ab}^+$ and $L_{ab}^-$
of $\Sigma$ considered as a submanifold in $M^+$ and in $M^-$ 
respectively are, in general, different. Together with the conditions
(\ref{mdelta}) this defines a well-posed spectral problem for an
operator of Laplace type. In this case there is a power-law
asymptotic expansion for the heat trace (no $\log$-terms)
with locally computable coefficients. First several coefficients
have been calculated by Bordag and Vassilevich \cite{Bordag:1999ed},
Moss \cite{Moss:2000gv}, Gilkey, Kirsten and Vassilevich \cite{Gilkey:2001mj}. 

It is very well known \cite{Albeverio}
that the conditions (\ref{mdelta}) are not the
the most
general matching conditions which can be defined on a surface. 
In general, boundary values of a function and of its' normal
derivatives are related by a $2\times 2$ {\it transfer} matrix:
\begin{equation}
0=
     \left(\begin{array}{cc}
          \nabla_{\nu^+}^++S^{++},\qquad&S^{+-}\\
          S^{-+},\qquad&\nabla_{\nu^-}^-+S^{--}\end{array}\right)
      \left(\begin{array}{l}\phi^+\\\phi^-\end{array}\right)
\bigg|_\Sigma\,.\label{BBT}
\end{equation}
Note, that the transfer conditions (\ref{BBT}) do not assume
identification of $\phi^+$ and $\phi^-$ on $\Sigma$. In other
words, there is no ad hoc relation between the restrictions
of the vector bundles $V^+|_\Sigma$ and $V^-|_\Sigma$. We can even
consider the situation when we have ${\rm dim}V^+\ne{\rm dim}V^-$,
i.e. the fields on $M^-$ and $M^-$ can have different structures
with respect to space-time and internal symmetries. $S^{\pm\pm}$
are some matrix valued functions on $\Sigma$ (one can even consider
the case when they are differential operators).
The heat trace asymptotics for the transfer problem have
been analysed by Gilkey, Kirsten and Vassilevich \cite{Gilkey:2002nv}.

Note, that although the conditions (\ref{mdelta}) can be obtained
from (\ref{BBT}) as a limiting case, taking asymptotic expansion of
the heat trace does not commute with this limit.

A rather interesting generalisation of the constructions considered
in this section consists in lifting the assumption that the metric
(or the leading symbol of the operator) is continuous across $\Sigma$.
Physically this means that the speed of light jumps on the interface
surface $\Sigma$, like in the case of a dielectric body immersed
in the vacuum. This modification, of course, complicates the
problem considerably. Therefore, except for a couple of particular
case calculations by Bordag et al \cite{Bordag:1998vs,Bordag:2001zj}
very little is known about behaviour of the spectral functions in this case.

In general, relations between dynamics in the volume and on the
boundary or on an interface is of much interest. Examples include
the boundary state dynamics in solid state physics and in strings,
the "near horizon" dynamics in physics of black holes. Celebrated
AdS/CFT correspondence principle 
\cite{Maldacena:1997re,Gubser:1998bc,Witten:1998qj} 
which states that
certain correlation functions in {\it quantum} conformal field
theory (CFT) on the boundary can be calculated by from {\it
classical} supergravity theory in the "volume" of the Anti-de
Sitter (AdS) manifold is also an example of such relations. In
this respect people sometimes refer to the so-called holographic
principle which is usually attributed to 't Hooft. Since no strict
formulation of this principle exists, it is usually understood as
any possibility to make statements about physics in the volume by
looking at the boundary. Here we have to remember again the
results on relations between spectral invariants of various
various differential operators acting on a manifold and on its´
boundary reported on this workshop.
\subsection{Supersymmetry}\label{ssusy}
Symmetry principles are the guiding rules of contemporary theoretical
physics. Therefore, for a long time physicists tried to find a
symmetry which would involve both bosonic and fermionic fields.
The crucial difference between bosons and fermion is that the former
obey a commutator algebra, while the latter an anti-commutator one.
This property introduces a natural grading both in the space of the
fields and in the symmetry group. An object which has to replace the
space time in the case of supersymmetric theories is the supermanifold
(see \cite{DeWitt:cy} for details). Locally a supermanifold
looks as a usual manifold with additional Grassmann coordinates 
$\vartheta^\alpha$ which satisfy the relation
$\vartheta^{\alpha} \vartheta^\beta = -\vartheta^{\beta} \vartheta^\alpha$
yielding $(\vartheta^\alpha)^2=0$. Functions on the supermanifold 
(``superfields'') are defined locally through a Taylor expansion in
$\vartheta$:
\begin{equation}
\Phi (x,\vartheta )=\phi (x) + \phi_\alpha (x) \vartheta^\alpha +
\phi_{\alpha\beta}(x) \vartheta^\alpha \vartheta^\beta +\dots
\label{supf}
\end{equation}
Since $(\vartheta^\alpha)^2=0$ the expansion (\ref{supf}) contains a
finite number of terms. One can also introduce an integration
and a differential structure on supermanifolds. Consequently, one
can define classical actions, classical field theories, and
develop a quantisation of these theories. This procedure gives
rise to some natural operators. 
As before, quantum effective action is defined by spectral functions
of these operators. Of course, these problems can be addressed locally
by using the expansions in the components (of the type (\ref{supf}).
However, a much nicer approach would consists in working directly
on a supermanifold without referring to a particular (super)coordinate
system. A modern survey on superanalysis can be found in 
the monograph by Khrennikov \cite{Khren}, 
which also contains a long list of open
problems.

I conclude this short section by noting that the enormous interest
to supersymmetry  in theoretical physics
is caused mostly by aesthetic reasons, and also because supersymmetric
theories are mathematically better behaved and, sometimes, even perturbatively
finite. 
So far there are no experimental
evidences in favour of supersymmetry.
\subsection{Non-commutative field theories}\label{snon}
Recent years much attention was attracted to non-commutative
field theories (see reviews by Douglas and Nekrasov \cite{Douglas:2001ba}
and by Szabo \cite{Szabo:2001kg}). 
Initially non-commutativity appeared in the framework of the
deformation quantisation approach (which means deformations of
symplectic structures in the mathematical language) of Bayen et al
\cite{Bayen:1977ha}
(for a  historical overview see \cite{Zachos:2001ux}).
Modern interest to the topic has been boosted by applications to
the solid state physics (quantum Hall effect) and to string theory
(cf.\ sec.\ \ref{sstring}).

The most rigorous way to introduce a non commutative manifold uses
the $C^*$-algebras \cite{Connes:ji}. For our purposes a somewhat 
simplified approach will be enough. In this approach one replaces
ordinary multiplication of the functions by the
Groenewold--Moyal product:
\begin{equation}
(f\star g)(x) = \exp \left( \frac i2 \theta^{\mu\nu} 
\partial_\mu^y \partial_\nu^x \right) f(y)\, g(x)\vert_{y=x}\,.
\label{GMpro}
\end{equation}
This product as it stays is defined for smooth functions only so that
it should be understood through Fourier series. 
The formula (\ref{GMpro}) is defined
with respect to a coordinate system. We shall consider a torus where
a suitable global coordinate system exists. It is easy to see that
\begin{equation}
x^\mu \star x^\nu - x^\nu \star x^\mu = i\theta^{\mu\nu} \label{scom}
\end{equation}
so that the associative product (\ref{GMpro}) reproduces the string
commutator (\ref{nccoor}).

It seems natural (and it is also useful for non-commutative QFT) to
study spectral properties of a generalisation of the Laplace type
operator:
\begin{equation}
D\phi = -\left( \delta^{\mu\nu} \partial_\mu \partial_\nu + a^\mu
\partial_\mu + b \right) \star \phi \,. \label{oper}
\end{equation}
Asymptotics the heat trace for
(\ref{oper}) were considered on a non-commutative torus by Vassilevich 
\cite{Vassilevich:2003yz}. These results were extended to non-commutative
$\mathbb{R}^n$ by Gayral and Iochum \cite{Gayral:2004ww}.
The main result is amazingly simple. There exists
a full power-law asymptotic expansion for the heat trace, and the 
coefficients are integrals of universal free polynomials of $a$, $b$ and their
derivatives evaluated with the non-commutative product. This also
shows that the heat trace coefficients have the form a ``typical''
classical
action for non-commutative field theories. In this respect, one
remembers the so called spectral action principle 
of Chamseddine and Connes \cite{Chamseddine:1996zu}
which suggests to use the heat trace asymptotics on non-commutative
manifolds to construct classical actions (this procedure is in a sense
``reverse'' of the renormalization). 

Here a warning is in order. One cannot expect that generalisations to
the non-commutative case will always be straightforward. Some operators
also appearing in the context of non-commutative QFT have quite
unusual spectral properties found by Vassilevich and Yurov
\cite{Vassilevich:2003he}.

It does not make much sense to give a list of open problems since
almost all problems at the interface of non-commutative geometry and
spectral geometry are ``open''. This seems to be a wonderful field
of research.

\end{document}